# Computer Assisted Access to Justice via Formal Jurisprudence Modeling

Michael Bar-Sinai
Ben-Gurion University of the Negev
Be'er-Sheva, Israel
barsinam@post.bgu.ac.il

Michal Tadjer
Worker's Hotline NGO
Tel Aviv-Jaffa, Israel

Mor Vilozni
CodeWorth.io
Tel Aviv-Jaffa, Israel

## ABSTRACT

This paper discusses an internet-based system for enabling people to self-assess their legal rights in a given situation, and a development methodology for such systems. The assessment process is based on a formal model of the relevant jurisprudence, exposed to the user through an interview. The model consists of a multi-dimensional space whose dimensions represent orthogonal jurisprudence aspects, and a decision graph that guides the user through that space. Self-assessment systems can revolutionize the way legal aid organizations help their clients, as they allow these organizations to deliver personalized help at internet scales. The proposed approach is validated through an implementation of a model for workers' rights when their employment ends. This model, describing Israeli law and developed in cooperation with a worker rights NGO, was ratified by external experts as accurate enough to be useful in real cases.

## 1 INTRODUCTION

Lack of knowledge of one's legal rights can lead to one's losing benefits and entitlements they deserve by law. This problem is aggravated when there are significant differences in access to knowledge between participants of a given dispute. Such differences are common in relations between low-income workers and their employers. In these cases, the workers are often members of disadvantaged groups and may not even speak the language at which the legal process is conducted, while the employers can afford professional legal services. Recent changes in the employment market, such as globalization and quasi-employment models used by companies such as Uber, make these worker-employer relations even less balanced.

The work presented here aims to use a computerized tool to help balance this inherent inequality between workers and employers, by providing workers with access to an online, interactive interview. This interview allows workers to assess their rights in certain situations. The interview is based on a formal model, which takes into account not only the law, but also regulations, prior rulings, and customary law.

The proposed approach is validated by implementing a model describing worker's rights when their employment ends, under Israeli jurisprudence. This model was created in collaboration with Kav LaOved (Worker's Hotline) a non-governmental organization (NGO) that protects worker's rights in Israel. It was ratified by experts outside of the development team as sufficiently accurate to be useful in real cases.

The main contributions of the work presented here are:

- Design and implementation of an interactive, online system for advising low-income workers when their employment ends. System output was validated by external experts as useful in real cases.
- Development methodology for creating models in similar legal fields.
- Application of the PolicyModels language and tool set [3], originally developed for the dataset privacy domain, to a legal rights related domain (specifically, labor law).
- Additions to the PolicyModels language and tool set (specifically: value inference, improved localizations, new language constructs).
- Presentation and discussion of design considerations for domain specific languages (DSLs) that focus on the legal field.

The rest of this paper is organized as follows: Section 2 provides required background on aid in the context of social services, such as those provided by government agencies and NGOs similar to Kav LaOved. Section 3 presents the PolicyModels language, and discusses some of its usages and design considerations. Section 4 presents the challenges in engineering real-world formal jurisprudence models, and proposes a methodology for developing them. Section 5 presents a case study of creating a real-world jurisprudence model using the proposed methodology. Sections 6 and 7 discuss related and future work, respectively. Section 8 concludes.

## 2 BACKGROUND: VIOLATIONS AND REMEDIES

This section presents the problem domain in which our system operates, including some theory and information on the low-income employment market.

A person whose rights have been violated, and wants to remedy the situation, has to go through three stages, as proposed by Felstiner et al. [5]:

(1) *Naming:* Coming to the realization that a violation had happened, and finding its legal definition (name).
(2) *Blaming:* Understanding who is to blame for the violation, and making the decision to confront them.
(3) *Claiming:* Asking the blamed entity to remedy the situation. If the request is turned down (wholly, or in part), the person may decide to transform the claim into a legal dispute.

An internet-based legal rights self-assessment system, such as the one presented here, can help people go through the first two stages of this process, and can help automate parts of the third. Aiding people through the *naming* and *blaming* stages is important for two main reasons. First, they are prone to errors: a perceived violation might not be a real violation, or the blamed entity might not be the correct one. Second, they are time-sensitive: our experience shows that chances of counteracting a violation and restoring the situation to its previous state (such as in an illegal dismissal) are greater when the first two stages of the process are done quickly after the violation.

At present, aid to the first two stages is provided by text-based self help web sites who attempt to make the law accessible, and by volunteers, who can provide personalized help through phone hot lines, email, internet-based chats, or meetings. However, the number of volunteers is limited, and the content in text-based websites is not personalized, which requires users to sift through a lot of legal material, and to decide on their own what parts are relevant to their case. While governments provide some aid [14], most of these websites and volunteer services are provided by NGOs.

Kav LaOved – Worker's Hotline (KLO)[1] is an NGO defending workers' rights in Israel. Established in 1991, KLO works both at the individual and the legislative levels. Individually, KLO helps about 55,000 workers each year, irrespective of nationality, religion, work sector, and legal status. In 2018, KLO returned approximately 10 million USD in illegally withheld income and benefits, to thousands of workers. At the legislative level, KLO works with legislators and other NGOs to promote pro-worker initiatives, and resists discriminatory laws and regulations by various means, including appeals to Israel's High Court of Justice.

KLO volunteers help workers through various channels: a phone hotline, email, Facebook page, and in-person meetings. Internet-based interaction (especially via social media) is now the primary source of legal aid – for example, KLO's migrant care giver workers Facebook page has over 48,000 followers (about two-thirds of the migrant workers in this sector in Israel). While internet-based communication is increasing, experience at KLO shows that workers still prefer in-person meetings when they face delicate situations such as employment termination. As the numbers of volunteers is limited (about 130), queueing for these meeting often takes hours.

The increasing strain on KLO's staff, combined with internet access becoming almost ubiquitous and the importance of aiding workers through the first two stages defined by Felstiner et al. in a timely fashion, are the backdrop for our proposed solution: an internet-based interview for self-assessing worker's rights and duties in a given situation.

In addition to helping workers, such interviews can guide volunteers, who are often not professionally trained legal experts. Moreover, these interviews can help employers as well.

Employers are often the stronger side in employer-employee disputes, but this is not always the case. Consider a person whose medical condition requires the constant aid of a care giver. Under Israeli law, the care giver is employed by said person, who might be struggling financially herself. If said person moves to a nursing home or dies, the care giver job is terminated, and she is entitled to severance pay based on the duration of her employment. If the employer or her mourning relatives have not prepared ahead of time, this sudden payment is likely to strain them financially. Such interviews can help them plan ahead, as they can know in advance what are the benefits the care giver will be entitled to.

We based our implementation on the PolicyModels language and tool set [3], which takes a model-based approach for assessing properties of a legal situation. As part of the work presented here, we have added new constructs to the language, and expanded its tool set.

## 3 POLICYMODELS: LANGUAGE AND TOOLS

PolicyModels [3] is a modeling language geared towards legal or legal-like domains. The term "legal like" here refers to domains with a set of rules intended for humans, that might not be laws in the strict legal sense. For example, PolicyModels is being used to evaluate scholarly systems with regards to academic data sharing guidelines[2]. The main purpose of PolicyModels models is to conduct interactive interviews which allow a layperson to assess the properties of a given case, with regards to the modeled policy. Being formal descriptions, these models can also be used to perform formal policy analyses, such as creating visualizations, or answering queries. For example, a model can be used to find all situations where certain properties hold (e.g. all cases where a worker job is terminated, AND she is not eligible for unemployment benefits).

A policy model consists of a policy space, a decision graph, automatic inferrers definitions, and textual localizations. We will now describe each of these components. This section provides a high-level description of the language and its design, and does not aim to be an exhaustive reference[3].

### 3.1 Policy Spaces

A *policy space* is a discrete space, containing all possible situations under the modeled policy – each point describes a unique situation. Policy space dimensions are ordinal, each describing a single aspect of the situation with regards to the modeled jurisprudence. In each dimension, coordinates represents possible values of that aspect. Figure 1 shows a policy space describing two legal aspects of job termination: worker age group, and process fairness. Coordinates are ordered to allow formal definition of sub-spaces such as "workers at or above work force age", or "workers at the pension age whose termination process was flowed or worse". This allows well-defined entitlement allocation, e.g. to workers at or above the voluntary pension age.

During computation, PolicyModels maintains a location in the model's policy space, representing the properties of the case being evaluated. This location is updated as the computation advances. Notably, for each dimension, these updates can only move the location upwards. This provides a sensible composition of the effects of various model parts.

For example, suppose part *a* of a model examines one set of legal aspects of a given employment termination case, and arrives

---

[1] https://www.kavlaoved.org.il/

[2] This work, done in collaboration with Force11.org, is described at https://www.force11.org/group/scholarly-commons-working-group/wp3decision-trees

[3] An exhaustive documentation, including language reference and tutorials, are available online at http://datatagginglibrary.readthedocs.io/

```
Root: consists of ProcessFairness, AgeGroup.
ProcessFairness: one of ok, flawed, illegal.
AgeGroup: one of under21, workForce,
              voluntaryPension, pension.
```

**Figure 1: A two-dimensional policy space describing worker age group, and process fairness (above), and the PolicyModels code defining it (below). The order of coordinates in a dimension can be used to determine which situations are more severe than others. For example, the black star represents a more severe situation than the circle.**

at the conclusion that the dismissal was flawed (e.g., the prior notice period requirement was not met). Thus, it moves the case coordinates to flawed along the ProcessFairness dimension. Part *b* of the model examines another set of legal aspects, according to which the dismissal was sound (e.g. the severance pay was accurate). Thus, it moves the case coordinates, along the ProcessFairness dimension, to Fair. As a whole, the layoff process in this case was flawed; we want the case policy space location to reflect this, regardless of the order in which parts *a* and *b* were executed.

The accuracy in which a policy space describes a given situations increases with its number of dimensions. However, more dimensions require the user to provide more information, making the model less usable. An efficient policy space contains only dimensions that describe aspects of a situation that affect the case under the modeled jurisprudence. The policy space describing worker's rights when their employment ends consists of 74 dimensions. One of these dimensions describes whether the worker is handicapped or not. However, this policy space does not contain dimensions describing that handicap, since – under Israeli Jurisprudence – these details do not affect worker's rights in this situation.

As in many other modeling domains, the decision which aspects to leave out is equally important as the decision what to include. The model has to be as simple as possible, but not simplistic.

## 3.2 Decision Graph

A decision graph describes a synergetic computation, where the computer and the user collaborate in order to find a location for a given case in the model's policy space (typically, the case the user is facing). The computer deals with the well defined parts of this process, such as maintaining the current case coordinates, or deciding where to move next in the graph, according to input from the user. The user decides on the softer questions, such as whether a "significant deterioration of employment conditions" was involved in the decision to quit one's job. This division of labor allows PolicyModels to overcome a main challenge of algorithmic treatment of legal cases, since is leaves the "soft", human-related decision to humans. Figure 2 shows a small part of the decision graph of our end of employment model.

Decision graphs are composed of nodes of various types. Each type of node is associated with a specific action. This action is performed by the computer when a computation arrives at the node. These actions can be asking the user a question ([ask] nodes), updating the case location coordinates ([set] nodes), invoking other parts of the graph ([call]s), etc. This level of indirection between user answers and updates to the case coordinates allows model developers to ask questions in a user-friendly manner, while using proper professional terms at the coordinate level.

As a whole, this synergetic computation process can be likened to a conversation between a vehicle owner and a mechanic in a garage: the mechanic asks questions that the vehicle owner understands; for example, "do you hear knocks from the engine at high speeds?" In accordance with the answers, the mechanic makes a note for himself of whether to check the spark plugs, the engine head gasket, or the timing belt -— terms that are too obscure for the average vehicle owner to provide useful information on.

A given policy space may have multiple feasible decision graphs. Specifically, a decision graph for a labor law specialist should be different than a decision graph for a layperson. From PolicyModels' point of view, the important part is not the decision process, but the final coordinate said process arrives at. This coordinate is used to present the interviewee with her legal status, rights, and recommendations. It can also be used by other systems that integrate with PolicyModels to provide additional help, e.g. towards the *claiming* stage.

Decision graphs can become rather large – our end of employment model presented here contains 251 nodes. In order to ease working with such graphs, PolicyModels defines nodes for dividing the graph structurally and semantically (see Subsection 3.6). Additionally, a graph definition can span multiple files.

## 3.3 Automatic Value Inference

Value inferrers, a new addition to the language, automatically update case location coordinate of one dimension based on its coordinates on other dimensions. They provide a declarative way of implementing statements such as "a person at or above working age, whose job was terminated in a flawed-or-worse process, is eligible to participate in a specific aid program".

Value inference in PolicyModels is based on the fact that every policy space location $l$ implicitly defines two sub spaces. A *compliance space*, which contains $l$ and all the locations whose coordinates

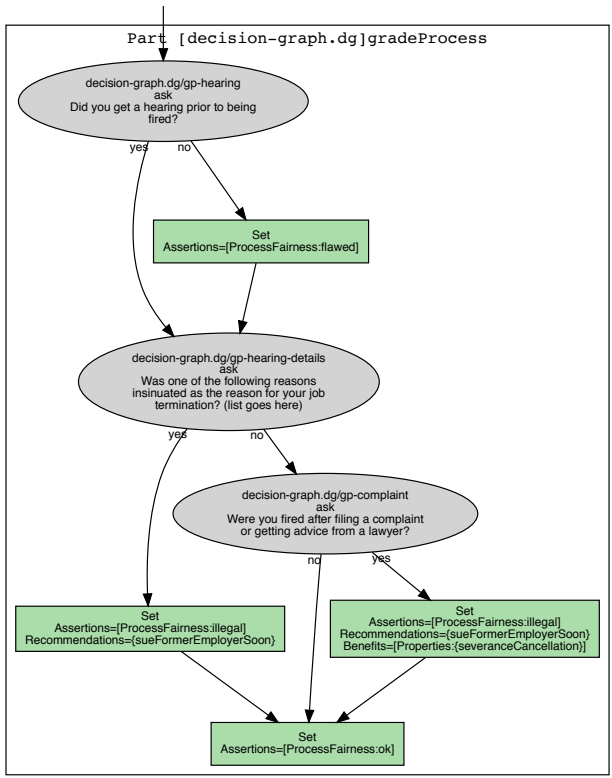

```
[>gp-hearing< ask:
  {text: Did you get a hearing prior to being fired?}
  {answers:
    {no:
      [set: ProcessFairness=flawed]}}]
[>gp-hearing-details< ask:
  {text: Was one of the following reasons insinuated
         as the reason for your job termination?
  {answers:
    {yes:
      [set: ProcessFairness=illegal;
            Recommendations+=sueFormerEmployerSoon]}
    {no:
      [>gp-complaint< ask:
        {text: Were you fired after filing a
               complaint or getting advice from a
               lawyer?}
        {answers:
          {yes:
            [set: ProcessFairness=illegal;
                  Recommendations+=sueFormerEmployerSoon;
                  Properties+=severanceCancellation]
          }}]}}]
[set: ProcessFairness=ok]
```

**Figure 2: Code and diagram of a decision graph for evaluating fairness of a job termination case. Shown graph contains nodes for prompting the user for details, and for updating case coordinates. As case coordinates only move upwards during computation, the last [set] node has no effect if prior nodes have marked the case as flawed or illegal. Diagram created by PolicyModels, using GraphViz [6].**

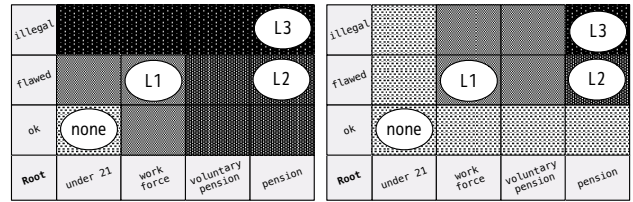

```
[Plan: support
  [AgeGroup=under21;  ProcessFairness=ok      -> None]
  [AgeGroup=workForce; ProcessFairness=flawed  -> L1  ]
  [AgeGroup=pension;   ProcessFairness=flawed  -> L2  ]
  [AgeGroup=pension;   ProcessFairness=illegal -> L3  ]
]
```

**Figure 3: Value inferrer for the Plan slot, inferring what plan a person is eligible for, based on her age and the fairness of their dismissal process. There are two types of value inferrers, differing on how they treat locations that do not explicitly appear in their definitions.** *Support* **inferrers (left), set the slot value at such locations by looking at the value defined for the closest point further from the space origin.** *Comply* **inferrers (right), use the closest defined location closer to the space origin.**

at each dimension are equal or bigger, and a *support space*, which contains $l$ and all the locations whose coordinates are equal or smaller[4]. To define a value inferrer, model developers list a series of locations in the policy space, and a value of the inferred slot at each one. These locations must form a series, where point $l_n$ is located at point $l_{n-1}$'s compliance space (that is, has at least one dimension with a bigger coordinate, and no dimensions with a smaller coordinate). This definition allows PolicyModels to set the inferred slot value at every location in the policy space, based either on the definition locations' compliance spaces, or on their support spaces. Figure 3 shows an example of both types of inference, as well as a sample value inferrer definition.

During computation, whenever PolicyModels updates the case location coordinates, it applies all model's value inferrers to further update that location. It keeps applying the inferrers until the location does not move. Since the location can only move upwards in each dimension, this process has to terminate.

Strictly speaking, value inferrers do not add expressive power to PolicyModels, as they can be implemented by manually adding conditional logic nodes after each [set]. From a software engineering standpoint, however, they make the models more readable and less prone to errors. Note that when using the manual alternative, missing single [set] node would introduce a bug to the model.

### 3.4 Localizations

The policy model parts described thus far contain mostly formal definitions, and very little text. The human-readable parts of the model are kept separately, in localization packages. Each localization package contains texts for questions defined in the model's decision graph, translated metadata (title, about text, etc.), and multi-level textual description of policy space entities: dimensions and coordinates. Each policy space entity is described in three levels: its name, a short sentence used as a tooltip, and a long explanation that may also include links, tables, and images. This multi level description is required since policy space entities often describe legal or technical terms and recommendations that laypeople are unfamiliar with, and thus may find intimidating.

---
[4]For a detailed discussion about these spaces, including the reason for their names, the reader is referred to Bar-Sinai et al. [3].

A single policy model can contain multiple localization packages. As is the case with many software localization technologies, creating a localization package requires very little technical knowledge. Thus, a single model can be translated to multiple languages. This is important for organizations such as KLO, that aid people from multiple international backgrounds who often do not read English readily.

### 3.5 Software Engineering a Using PolicyModels

The PolicyModels technology stack is designed using a model-based software engineering approach [13]. A PolicyModels runtime engine (a Java library) is embedded in a larger application. During an interview, the application feeds information (such as user input) to the runtime engine, and is being notified when these cause changes to the model, such as an update in the case location coordinates. At the time of this writing, the PolicyModels runtime is embedded in two applications: a web application for conducting on-line interviews, written using Scala[5] and Play Framework[6], and a command-line application for processing, debugging and testing models, written in Java.

### 3.6 On Creating a Legal-Oriented DSL

PolicyModels defines two domain specific languages (DSLs), one for declaring policy spaces, and one for creating decision graphs (we feel the value inference syntax is too simple to be called "a language"). This subsection presents some of the design considerations behind these languages.

PolicyModels is intended to be used by domain experts, not by computer science graduates. Thus, PolicyModels' DSLs cannot rely on the programmer being familiar with common programming concepts such loops, branches, procedures, or classes. Moreover, a policy model is not a program, but a model with execution semantics; it does not contain imperative instructions to a computer, but rather describes what could be done given a series of user inputs. This limits the model's expressive power (PolicyModels is not Turing complete) but facilitates formal analysis, which is a desirable trait.

A policy space consists of numerous dimensions. To allow model developers to properly arrange these dimensions, policy spaces are defined using slots of various types. An atomic slot is a direct representation of a dimension: its definition contains the dimension coordinates, in order. An aggregate slot defines a set (in the mathematical sense, not the data structural sense), whose possible members come from a predetermined list. From a theoretical point of view, aggregate slots are syntactic sugar for defining multiple atomic slots whose coordinates are false, true (in that order). A compound slot groups other slots together. These slots only matter for model engineering – they do not add dimensions to the defined policy space.

To provide basic human-readable text during early stages of model development, all definitions of policy space entities can contain a textual remark. This remark is also used in creating new localization packages. See bottom of Figure 1 for a sample definition of a policy space.

Decision graphs are defined such that they resembles interview instructions, in the same way a domain expert would instruct a volunteer on how to perform an interview with a worker. Thus, node types are in the imperative: [ask], [set], and [consider] (PolicyModels' version of the classic switch control flow construct). To allow localizations, all nodes can have ids, so external data structures can reference them in a stable manner.

PolicyModels' decision graph DSL does not have a concept of variables. During runtime, the only value that exists is the current location in the policy space. If the need for a helper variable arises (indeed a common case), model developers can define helper slots – not all policy space's dimensions need to have external semantics.

One concept we were not able to keep out of the decision graph DSL is the *call stack*. Some interview sequences have to be performed in multiple scenarios, and supporting this requirement by re-using parts of the graph through a calling mechanism seemed to be the simplest way. The division of a decision graph to parts that are easy to work with domain-wise and engineering-wise proved to be challenging, as these aspects require different divisions.

PolicyModels supports two ways of dividing a decision graph: for graph engineering reasons (using [part]), and thematically (using [section]). Our experience shows that these divisions are different: engineering considerations often require code division supporting easier navigation and reuse, while legal considerations require structuring the code by subject.

Using a [section] node states that, e.g., "this part of the graph deals with assessing the legal status of the interviewee". [section]s cannot be invoked from other parts or the code. In this sense, [section] is similar to code blocks in Algol-like languages (such as C and Java), where code blocks group statements and introduce a new scope, but do not add any semantics to what is being done. While code blocks are not commonly used, we find that [section]s are used quite a lot in our code. This might be due to the synergistic nature of PolicyModels' computation, which benefits from explaining to the user what she and the computer are currently working on.

A [part] node (marked in code by [--> ... --]) defines a re-usable decision graph, similar to a procedure in a regular programming language. It can be invoked from multiple places in the graph, by using a [call] node.

The distinction between [section] and [part] is new. We have arrived at it after a few failed designs iterations, described below. These iterations were done in parallel with, and informed by, the creation of increasingly complex models.

*Intentionally Primordial Stage.* Our initial approach was to start with a basic call mechanism similar to goto, and then add control structures based on common usage patterns. At this stage, a [call] could invoke any type of node, and would place it on the call stack.

---
[5]https://www.scala-lang.org
[6]https://www.playframework.com

When an interview process arrived at an [end], the call stack was popped and the computation continued from the node after the popped [call].

*Division to [section]s.* The [section] node was added to allow model developers to thematically group parts of a decision graph under a specific title. Based on our experience, where most reusable parts revolve around the same domain-related subject, and that decision graphs where any node could be [call]ed were poorly structured, we decided to use [section]s for structural grouping as well. As a result, a [section] could be executed in two possible ways: being invoked from a [call] node, or as part of the usual graph traversal (i.e. being the next node of a node that was executed). This makes the execution semantics of [section]s ambiguous, as [end] nodes in them behave differently according the way the [section] was entered. We were aware of this, but hoped the limited language complexity would prevent the semantic issue from being manifested. This was not the case, mainly because PolicyModels is a modeling language, and the introduced ambiguity made its analysis tools, such as visualizations, much less usable. We consider this a failed experiment, and note (yet again) that language semantics have to be kept clean, always.

*Theme and structure, separated.* This is the current status of the language, where thematic grouping is done using [section], and structural grouping is done using [part]. As in previous stages, the [end] node terminates the execution of the current [part]. A new node, [continue], terminates the execution of the current [section]; after arriving at a [continue] node, the computation proceeds to the next node after the current [section]. While this stage contains the most concepts (4 node types), we find it useful, readable, and semantically sound.

The PolicyModels DSLs support top-down design approach at the language level, by using "TODO" constructs. Slots defined as TODO can be localized and grouped under compound slots. [todo] nodes act as placeholders for parts of the decision graph that should be implemented at later stages. We found these constructs useful during the creation of large models, as discussed in the next section.

## 4 ENGINEERING A JURISPRUDENCE MODEL

This section discusses the process of creating a real-world policy model, based on our experience in creating the end-of-employment model and others. As policy models are very similar to small computer programs, the proposed process is based on common software engineering methodologies. We do not claim that this is the best possible method; rather, our intention is to offer a sufficient method in order to allow others to start creating models and to initiate a discussion on the subject. We begin by discussing some unique challenges model development teams have to overcome, and then propose a development methodology.

Policy models are legal-technological hybrids. Thus, building a policy model for a certain legal field requires expertise in two areas: the legal field and the PolicyModels system. Thus, a model-building team will usually be made up of at least two experts from different backgrounds, who will not be familiar with the complementary field. It is important to note that the level of expertise necessary in each field is different. Legal experts are required to have deep understanding of the legal field, in addition to remaining up-to-date in it (for example, being familiar with recent rulings). In contrast, a person with basic training in computer programming can use PolicyModels after a relatively short amount of study. Our experience shows that computer science students are able to use PolicyModels after reading the training documents. Therefore, we estimate that a programmer with little experience can start building models after one day of self-teaching. Clearly, the programmer's efficiency will increase with experience.

Cultural differences between computer programmers and jurists are another challenge that must be bridged, preferably early in the work process. These cultural differences stem from the discursive and conceptual gaps between law and technology [4]. Unlike computer science, the legal field is not set up for unambiguous structures; it is no accident that legal decisions span dozens or hundreds of pages. The legal field is composed of primary legislation, directives (secondary legislation), case law that has been set out in court, and even procedures of government ministries. Legal interpretation of a person's situation requires clarifying information from his or her life events, giving them a legal headline.

Many computer programmers – authors with computer science background included – have difficulties coping with fields that have a range of contradictory opinions, such as the legal field; jurists, for their part, must become accustomed to thinking about legal situations in formal terms, such as the policy space and decision graphs.

To allow a model to answer which rights and obligations result from a certain personal situation (such as in the termination of employment of a migrant worker after a heart attack), jurists must be willing to let go of the distinction between what is set in law, and therefore is ranked higher, and what is determined by the Interior Ministry procedures, which have never undergone judicial review. The law presented in a policy model is simplified and adapted to PolicyModels' limitations; it cannot be intricate and hierarchical, as it is in court rulings, petitions, and lawsuits. However, it is this simplification that allows for the self-help process, and for the clear recommendations users can put to use.

### 4.1 Model Development Methodology

We now turn to our proposed development methodology, which is based on the iterative software development process [11]. This methodology creates a working model early in the process, which allows for earlier evaluation and error correction. Working models also help make the process and products tangible earlier, which is important for the motivation of the domain experts, who may not be used to seeing a software tool in its early, non-usable development stages.

The development process is made of two stages: setup and planning, and iterations. We detail them below.

*4.1.1 Team Setup and Project Planning.* At this stage the jurists introduce the technical team members to jurisprudence relevant to the field in question, and the technical members introduce the capabilities and limitations of PolicyModels to the jurists. Then, the team decides what parts of the jurisprudence can be modeled, and how to approach this modeling (e.g. where to start).

An implicit but important goal of this stage is to acknowledge the aforementioned cultural differences, so that it is easier for the team to bridge over them during subsequent stages.

*4.1.2 Development Iteration.* Similar to iterations in the classic iterative software development process, each iteration produces a working model that can be used by testers, reviewers, and users. We found that a top-down development direction, where each iteration increases model accuracy, works better than the bottom-up direction. This is, in part, due to the fact that PolicyModels languages support top-down development through specialized language constructs (see Subsection 3.6).

*Iteration phase 1: Planning.* At this stage, the team selects a sub-field of the field being modeled ("breadth"), and decides on the level of accuracy in which that sub-field will be modeled ("depth"). Normally, initial iterations will be broad and shallow, while later iterations will be narrow and deep. This also allows bringing external sub-domain experts for short and intensive consultation sessions, which are easier to schedule than numerous meetings over a long period. This is an important consideration, since the core modeling team may not have all the required legal expertise.

*Iteration phase 2: Implementation.* The development team surveys the sub-field, and updates the policy space and decision graph accordingly. Parts that are identified but whose detailed modeling is left to later iterations are marked as incomplete, using [todo] in the decision graph, and TODO slots in teh policy space. PolicyModels can create a report listing these parts, as well as unused dimensions and values in the policy space, so that the team can have a clear picture of what parts of the model are not implemented yet.

As mentioned above, PolicyModels can create diagrams of the decision graph and policy space. These diagrams are useful at this stage, as they allow non-technical team members, who might initially find code intimidating, to see that the model implementation accurately reflects their intensions.

*Iteration phase 3: Review/QA.* The model is tested by the core team. Testing is done by going through an interview, providing details of specific cases, and inspecting the resulting recommendations. PolicyModels server supports this stage in two ways. First, it allows collecting comments from interviewees. Comments are linked to specific model parts and a specific localization package, allowing them to refer to so both structural and textual defects. Second, it allows administrators to mark a model version as "private". Private versions are accessible only through sharable private links – they are not listed in the public pages. If a review from an external expert is required, the core team can send a sharable link to said expert, without making the model publicly available.

*Iteration phase 4: Release.* After the issues discovered at the previous stage are fixed (or marked for solving in subsequent iterations), the core team makes the new version publicly available by uploading it to a PolicyModels server, and marking the model version "public".

Our experience in developing several models shows that a core team of two people – a legal expert and a modeling expert – works well enough. A series of weekly work meetings proved to be a relatively effective way to build the model. The length of each of these meetings ranges between two and four hours, and sometimes even more – depending on the time constraints of team members, and their stamina.

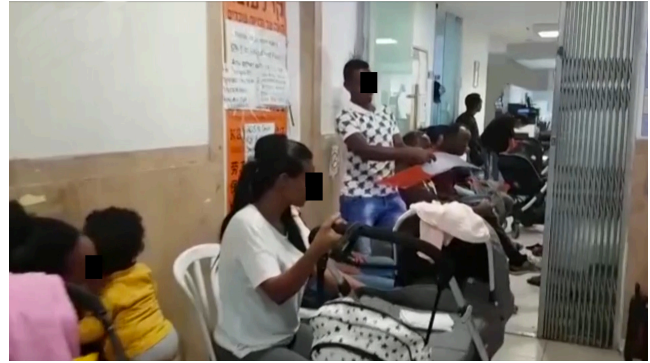

**Figure 4: Reception Hours at the Tel-Aviv Kav LaOved branch.**

## 5 CASE STUDY: WORKER RIGHTS AT END OF EMPLOYMENT UNDER ISRAELI LAW

We now describe a case study of developing a policy model covering the rights and duties of workers under Israeli law, developed by the authors. This model also provides recommendations to the worker, based on her status. The source code for the model, as well an interactive version of the interview, are available online[7]. We will start by describing the current status at KLO, which forms the backdrop for this project, and then describe the project and its progress.

KLO is an Israeli NGO dedicated to protecting workers and their rights. Established in 1991, KLO works mainly with low-income workers, who might be Israeli citizens, migrant workers, undocumented workers, palestinians, or members of an other disadvantaged group. Not all workers speak Hebrew or English. Some work at remote locations, which means consulting with a KLO volunteer in person requires them to take a day off. KLO operates phone hotlines, maintains Facebook pages based on work sector, and produces textual help leaflets. KLO also operates three branches where volunteers can meet and consult workers in person.

KLO's team consists of about 20 employees and 130 volunteers, who aid 55,000 workers every year. As can be expected, workers have to queue for hours in order to consult with a volunteer (see Figure 4). Most workers can access to the internet through their smartphones. This project sought to improve KLO's service and alleviate the pressure on its volunteers, by providing an online system that allows workers to deal with the simpler cases on their own. Figure 5 shows a mobile phone during an interview, and cards with QR codes that invite users to self-assess their rights when their employment ends.

Experience at KLO shows that people, particularly those who are members of marginalized populations, do not know how to describe what has happened to them using legal terms. Therefore, our model does not refer to job termination as "dismissal," "resignation" or "resignation under dismissal circumstances." Instead, it uses phrases such as "the work has ended", which anyone can acknowledge.

---

[7]Latest public version: https://klo-rights.codeworth.io/models/end-of-employment/latest.
Model source code: https://github.com/codeworth-gh/KLO

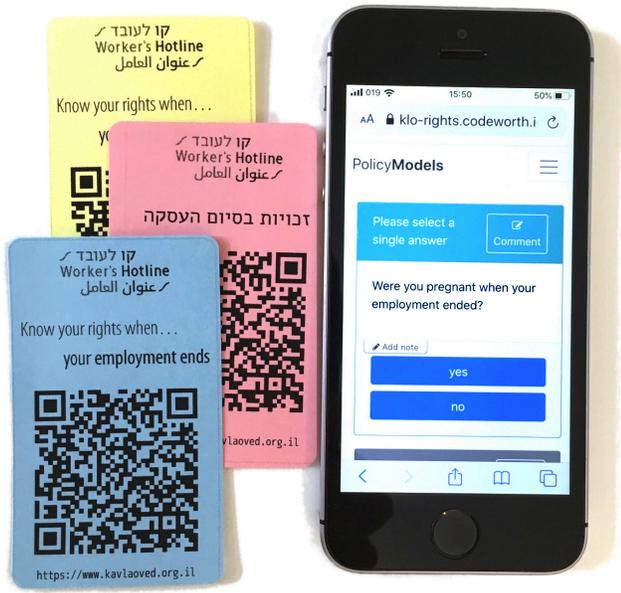

Figure 5: Most low income workers in Israel have internet access via their mobile phones. Thus, self-help interview web applications, such as PolicyModels Server pictured here, must support mobile devices. Cards with QR codes (left) can facilitate usage of internet-based self help systems, by advertising their existence, and helping users open them on their mobile phones.

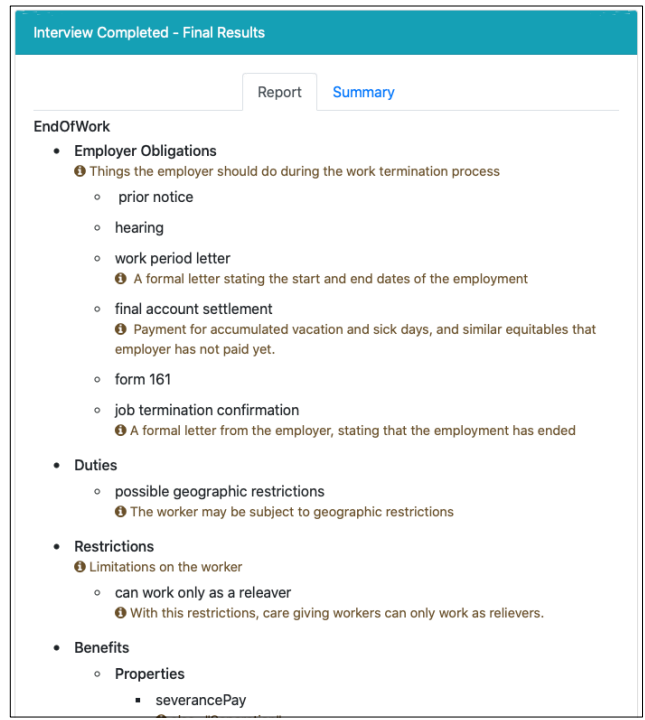

Figure 6: The final report shown to workers and the end of the end-of-employment interview. This report details what are the worker's entitlements, duties, and restrictions, as well as the employer's obligations to the worker. A recommendation section, listing actions the worker can consider doing (but does not have to), may also be present.

In accordance with the methodology presented in Section 4, our project started with a mutual introduction of the legal domains KLO works at, and the PolicyModels tool set. This was done in a 4 hour workshop, attended by the extended project team. At the workshop, the team decided to model the end-of-employment domain. This domain was chosen because KLO helps many workers in this area, and because it can be described using a policy space accurately enough.

After the initial workshop and the selection of the domain, a team of two members – a legal domain expert and a modeling expert – was formed. The team held work meetings on average 3 times a month for 9 months, occasionally consulting with external domain experts. An initial version of the interview was ratified by two KLO labor lawyers who were not part of the team, after 8 month of work.

The complete model covers various legal statues (undocumented, asylum seekers, palestinians with work permit, care giving work visas, agricultural work visas, and Israeli citizens). Interview outputs contain worker entitlements and duties, recommendations (action the worker can, but does not have to do), and the employer obligations to the worker (see Figure 6). This report can empower workers at highly vulnerable situations. Examples include:

- When a woman's employment is terminated while she is pregnant or during a parental leave, the system informs her that her dismissal might be illegal, and recommends contacting the commissioner for the Employment of Women Law at the Ministry of Labor, Social Affairs and Social Services.
- When a worker resigns because of sexual harassment, the model informs them that they have the same entitlements as if they were fired, even though the decision to terminate the employment was their own. Knowing that these benefits, which include severance pay, will be available to them if they decide to leave, can empower workers to resign from sexually abusive work places.
- When an employer shuts down a company and does not pay their employees' last salary citing financial reasons, the model recommends that the employees organize and file a bankruptcy request, which will allow them to get their pay[8].

---

[8]KLO have helped workers in a few cases where an employer avoided paying salaries by shutting down one company, claiming there was no money left, and then continuing business as usual by opening a new company and hiring new people. Having employees file for official bankruptcy is one way of combating this loophole.

## 6 RELATED WORK

A common format for internet-based self-help systems is a text-based site, organized around the "know your rights" theme. Example of such sites include ACLU's Know Your Rights[9], Civil Law Self Help Center[10], and Israel's Kol Zchut (literally: "every right")[11]. These sites are thoughtfully organized, support search, and successfully simplify complex legal terms. However, they leave the legal reasoning to their readers. Users are required to read though large amounts of text, some of which describes legal situations and rights that do not pertain to their individual cases. The personalized process offered by PolicyModels' interactive interviews improves on this situation in that the interview contains only relevant questions and answers, and the final report consists only of information relevant to the interviewee's specific case. In most cases, a worker in distress does not have the capacity to find the "needle in the haystack". This is evident from looking at typical reception hours at a KLO branch, where workers prefer the lengthy queues for volunteer consultation over reading the available information leaflets.

Another way for workers to get help is through interaction with other people in social networks and internet fora. While these systems can provide a sense of community (which is highly important in its own right), they rely on other users to supply legal advice. These users may not have the expertise and time required to provide good advice.

PolicyModels was originally developed as part of Datatags [16], a system for regulating the sharing and handling of scientific datasets containing sensitive information. In this context, it was used to create an interactive interview for recommending a datatag (and, thus, sharing and handling guidelines) based on answers from the dataset depositor. The interview, implemented using a policy model, aims to capture the relevant legal and technological knowledge required to assign a datatag to a dataset. Thus, the interviewee only has to know details pertaining to the dataset in question.

The Datatags use case can be paralleled to the use case presented here. The Israeli labor-related jurisprudence is equivalent to the US privacy laws and available storage technologies, and the worker's case is equivalent to the dataset. In both cases, the policy model captures domain knowledge, and interviewees – scientists or low-income workers – only have to provide the details of their own case in order to receive a useful answer.

PolicyModels was used to model unemployment benefits under Israel's National Insurance law [2]. Similar to the work presented here, the PolicyModels tool set was applied to the social policy domain. However, unlike the work presented here, only parts of the law were modeled, and the model did not take into account case law, regulations, and other information outside the written law text. This was a successful experiment, paving the road for the real-world example presented here.

PolicyModels is of course not the first system to offer help in legal areas through an interactive, on-line interview. TurboTax[12], which was one of the inspirations to PolicyModels, helps taxpayers in the United States to complete their personal tax forms.

The work presented here requires a legal domain human expert to interpret the law. Ghaisas et al. [7] offer an alternative, by using deep learning to disambiguate the law. The framework they present disambiguates regulatory texts by integrating and summarizing external textual sources, to allow software engineers to create software systems that follow the spirit of the law. They have demonstrated their Ambiguity Resolution Framework on the United States' Health Insurance Portability and Accountability Act (HIPAA). However, AI systems can have implicit biases, e.g. demonstrate racist behavior [1, 15]. Thus, using AI in socially sensitive contexts, such as the one presented here, requires caution. Nevertheless, such disambiguation techniques can aid model development teams – legal domain experts included – while interpreting the law.

## 7 FUTURE WORK

The work presented here addresses the first two stages of remedying a violation of a worker's rights (see Section 2): *Naming* and *Blaming*. Future work can build on interview results to help with the third stage: *Claiming*. This could be done as the interview result – a location in the model's policy space – has well-defined semantics, and can easily be communicated in a machine-readable way.

For example, suppose a given interview result contains an indication that the lay-off process they describe was flawed. These results can be used to create a formal letter the employee could send her employer, in order to claim compensation.

A similar technique can be used to implement benefit calculators. Consider the case of severance pay (under Israeli law). The amount due, and the way it is calculated, depends on the length of employment, its units (hourly, daily, or monthly), its extent (full vs. partial), and on other aspects. A severance pay calculator could use a PolicyModels interview to determine how the amount due will be calculated. After receiving the interview results, the calculator will know what additional data to request from the user in order to complete the calculation (e.g., average amount of hours worked in the last two months is only relevant for hourly employment).

Lengthy question sequences can be tiring, and so users may find it hard to complete an interview with a detailed model. Once a few models are in public use, it will be interesting to look into their usage statistics and examine their usability. Results may inform the design of decision graphs and the interview UI, so as to make detailed interviews less taxing.

Being a formal representation of a jurisprudence interpretation, policy models can be used as a tool for comparative analysis of law systems or interpretations. Querying these models can be used to analyze a law system, by identifying cases where certain properties hold. This technique can be used to answer questions such as "what are the situations in which a worker job was terminated, but she is not eligible for severance pay".

PolicyModels is released under an open-source license, like many other open source systems developed for aiding human rights centers and NGOs [8]. We hope this will facilitate the creation of a policy modeling community, helping additional disadvantaged communities that the developers relate to, or are members of. This hope is based on the fact that open source contributors' performance is positively affected by the impact of the software they develop, and its *meaningfulness* ("the perceived value of the task in relation to one's personal beliefs, specifically attitudes and values") [9]. Grassroots development communities have developed for similar civic needs, such as crisis mapping [12] and open government [14]. Knutas et al. [10] show that, even though these communities may struggle with engaging stakeholders, requirement specification, and product delivery orientation, they are able to deliver software systematically, and support its evolution.

---

[9] https://www.aclu.org/know-your-rights/
[10] https://www.civillawselfhelpcenter.org/
[11] https://www.kolzchut.org.il/
[12] https://turbotax.intuit.com/

## 8 CONCLUSION

In this paper, we demonstrated that model-based interviews can be used as legal self-help aids in the first two stages of receiving legal help: they can be used to name the harm done, and to identify the entity or institute that should remedy the situation.

The transition from law to an unambiguous and simple format is undoubtedly complex and not suitable for all legal fields. A successful modeling process requires choosing a specific situation, whose legal conclusions are relatively simple, and keeping in mind that this tool may not always provide the best treatment. Indeed, in cases where our model detects severe harm, such as sexual assault or exploitation, we refer the interviewees to the proper aid services. It is worth emphasizing that the very act of identifying the offense and locating the right institution to contact are part of the solution.

## ACKNOWLEDGEMENTS

This work was funded in part by grant CNS-1237235 from the National Science Foundation, and by a grant from the Israeli Innovation Authority's Technology in Government track. The authors thank Stephen Chong, Alexandra Wood, and Mercè Crosas of Harvard, Shevy Korzen of the Israeli Public Knowledge Workshop, and the extended staff at Kav LaOved.